\documentclass{emulateapj}  

\usepackage{amsmath}
\newcommand{\electrons}{e^{-}}
\newcommand{\photons}{\mathrm{photons}}
\newcommand{\hrmin}[2]{#1^{\mathrm{h}}#2^{\mathrm{m}}}
\newcommand{\ra}[3]{#1^{\mathrm{h}}#2^{\mathrm{m}}#3^{\mathrm{s}}}
\newcommand{\dec}[3]{#1^{\circ}#2{\arcmin}#3{\arcsec}}

\def\deg{\ifmmode^\circ\else$^\circ$\fi}
\def\Fscr{\ifmmode{\mathcal{F}}\else$\mathcal{F}$\fi}
\def\Ascr{\ifmmode{\mathcal{A}}\else$\mathcal{A}$\fi}
\def\spie{Proc. of the SPIE}

\makeatletter
\shorttitle{$y$ Band Sky Variability}
\shortauthors{High et al.}
\begin{document}

\title{Sky Variability in the $y$ Band at the LSST Site}

\author{F.\ William High, Christopher W.\ Stubbs \& Brian Stalder}

\affil{Department of Physics \\
  and \\
  Harvard-Smithsonian Center for Astrophysics \\
  Harvard University, Cambridge, MA 02138 USA}
\email{high@physics.harvard.edu}

\author{David Kirk Gilmore}
\affil{Kavli Institute for Particle Astrophysics and Cosmology\\
Stanford Linear Accelerator Center\\
Sand Hill Road, Palo Alto, CA 94025 USA}

\author{John L.\ Tonry}
\affil{Institute for Astronomy\\
University of Hawaii\\
2680 Woodlawn Drive, Honolulu, HI 96822 USA}

\begin{abstract}

  We have measured spatial and temporal variability in the $y$ band
  sky brightness over the course of four nights above Cerro Tololo
  near Cerro Pachon, Chile, the planned site for the Large Synoptic
  Survey Telescope (LSST).  Our wide-angle camera lens provided a
  $41\;\deg$ field of view and a $145\arcsec$ pixel scale.
  We minimized potential system throughput differences by deploying a
  deep depletion CCD and a filter that matches the proposed LSST $y_3$
  band ($970$ to $1030\;\mathrm{nm}$).  Images of the sky exhibited
  coherent wave structure, attributable to atmospheric gravity waves
  at $90\;\mathrm{km}$ altitude, creating 3\% to 4\% rms
  spatial sky flux variability on scales of about 2 degrees and
  larger.  Over the course of a full night the $y_3$ band additionally
  showed highly coherent temporal variability of up to a factor of 2
  in flux. We estimate the mean absolute sky level to be approximately
  $y_3=17.8\;\mathrm{mag (Vega)}$, or $y_3=18.3\;\mathrm{mag (AB)}$.
  While our observations were made through a $y_3$ filter, the
  relative sky brightness variability should hold for all proposed $y$ bands,
  whereas the absolute levels should more strongly depend on spectral
  response.  The spatial variability presents a challenge to
  wide-field cameras that require illumination correction strategies
  that make use of stacked sky flats.  The temporal variability
  may warrant an adaptive $y$ band imaging strategy for LSST, to take
  advantage of times when the sky is darkest.

\end{abstract}


\section{Introduction}
 
The Large Synoptic Survey Telescope (LSST) is a proposed wide-field
camera that is currently in the design and development phase
\citep[][hereafter LSST09]{LSST}. The project has designated the
telescope's site to be Cerro Pachon, Chile, on the same ridge that
hosts the Gemini South and SOAR telescopes. The LSST will survey the
sky in the Sloan Digital Sky Survey's $ugriz$ passbands, plus an
additional near infrared band, $y$, at wavelengths $\sim
990\;\mathrm{nm}$, only recently made possible by deep depletion CCD
technology that enhances quantum efficiency (QE) at the reddest
wavelengths \citep{OConnor}.  The Pan-STARRS \citep[][]{PS02} survey
also uses the $y$ band.

The designation of ``$y$'' may present confusion.  Infrared
astronomers use a band $Y$, typically with a central wavelength near
$1.03\micron$ \citep{Hillen02,Hew06}.  Their use of infrared
detectors, with a flatter response curve near a micron, yields a
passband with significantly different shape than the CCD-based
$y$. Our filter also bears no relation to the optical $y$ filter of
the Stromgren passbands, which is centered on $550\;\mathrm{nm}$. This
degeneracy in terminology is most unfortunate, but seems likely to
persist.

The sky brightness in the $y$ band has not been well characterized for
astronomical CCD applications.  Past work describing sky brightness at
observatories typically cite values for the $UBVRI$ bands, but not $y$
\citep{Patat03,KK07,Sanchez07}. The reddest band in the SDSS survey is
$z$, whose red edge is determined by the rapidly falling QE of the
SDSS detectors, which have very little sensitivity in the $y$ band
region.  The UKIRT survey presented sky measurements for the $Y$ band
\citep{Warren07}, but as explained before, 
the total transmission curve is significantly different than
that using a deep depletion CCD.

\begin{figure}
\plotone{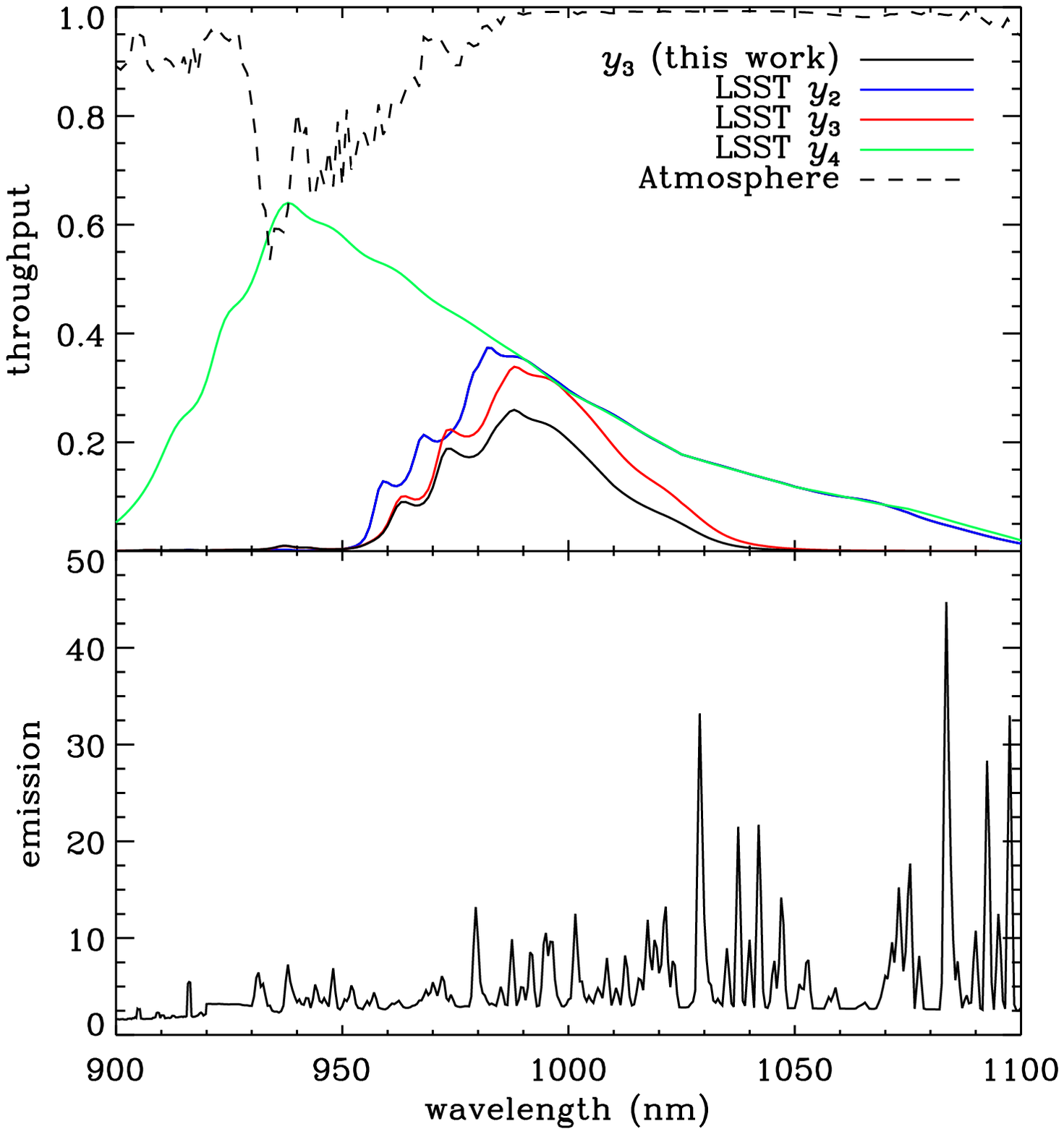}
\caption{The LSST $y$ band and its variants (top), and the sky
  emission spectrum (bottom).  The top panel shows the expected
  transmission spectrum for proposed $y$ band variants \citep{LSST} 
  including predicted optics and detector response,
  as well as the relative transmission that we measure for our apparatus (black
  solid line).  Our effective filter curve closely resembles the planned
  $y_3$ band.  All spectra exclude atmospheric absorption and
  scattering, which we plot separately as a dashed line. The
  bottom panel shows a typical sky spectrum in
  $\photons\;\mathrm{s}^{-1}\;\mathrm{m}^{-2}\;\mathrm{nm}^{-1}\;\mathrm{arcsec}^{-2}$,
  dominated by Meinel band OH emission lines \citep[taken
  from][]{Puxley}.  
}
\label{fig:Pixis_QE}
\end{figure}

Our aim has been to provide the first dedicated CCD measurements, at
the proposed LSST site, of the $y$ band sky background and
its variability over the course of a night.  While the exact character of the new $y$ band to be used for
LSST is under active consideration, we used a filter candidate called
$y_3$ (LSST09). We expect our qualitative findings to hold for any $y$
band variant under active consideration because the sky background in
all cases arises overwhelmingly from narrow OH emission lines.  Figure
\ref{fig:Pixis_QE} shows potential $y$ band transmission curves with
detector quantum efficiency (QE) and atmospheric response for both
LSST and our apparatus, along with the anticipated sky emission
spectrum.

In \S \ref{sec:nature} we briefly describe the known nature of sky
emission within the $y$ passband, covering past aeronomic (\S
\ref{sec:aero}) and astronomical (\S \ref{sec:astro}) measurements.
In \S \ref{sec:apparatus} we describe the dedicated instrument we
assembled for these measurements.  Section \ref{sec:observations} discusses
the observations, and the data reduction is outlined in \S
\ref{sec:reduction}.  Section \ref{sec:results} presents results on spatial
(\S \ref{sec:spatial}) and temporal (\S \ref{sec:temporal})
variability, and an absolute calibration (\S \ref{sec:absolute}), and
we conclude with \S \ref{sec:conclusions}.

\section{The Nature of $y$ Band Sky Background}
\label{sec:nature}

The background sky in the $y$ passband is dominated by Meinel band
emission due to hydroxyl \citep{bib:meineloh1,bib:meineloh2}, in a
physical process that has long been observed
\citep[eg][]{wiens,bib:nakamura} and modeled
\citep{rousselot,cosby,cosbyslanger} by atmospheric scientists
\citep[for a useful overview, see][]{marsh}.  Hydroxyl is well
stratified in the mesopause at $90\;\mathrm{km}$ in a
$10\;\mathrm{km}$ thick layer \citep{baker_stair,melo}.  Above the
ozone layer in the stratosphere, daytime UV light from the sun
dissociates $\mathrm{O}_2$ to make $\mathrm{O}+\mathrm{O}$, which
react with other $\mathrm{O}_2$ to produce $\mathrm{O}_3$.  UV light
also dissociates $\mathrm{O}_3$, completing the oxygen-ozone cycle.
$\mathrm{O}_3$ additionally reacts with $\mathrm{H}$ to produce
$\mathrm{O}_2$ and excited $\mathrm{OH}$, which then emits photons to
produce the airglow within the $y$ band.

As distinct from Rayleigh scattering, Meinel emission is known to vary
significantly over a number of spatial and temporal scales
\citep[eg][]{marsh}.  
The airglow varies with the tides, and is linked to the seasons and
the solar cycle.  Traveling and standing atmospheric gravity waves,
which are coherent density perturbations analogous to ocean waves, are
regularly observed by aeronomers
\citep[eg][]{taylor,garcia,bib:nakamura,li}, and give rise to a
variable background over a single, wide-field $y$ band CCD image.


\subsection{Aeronomic Measurements}
\label{sec:aero}

Aeronomers, like astronomers, have made extensive use of CCDs since
their invention, but also have not explored the $y$ passband region
with them in depth because the development of deep depletion is
recent.  CCDs have been used to measure sky emission lines in
isolation with Fabry-Perot interferometers, for the purpose of
monitoring atmospheric gravity waves and variability in species
temperature and emission rate \citep[eg, the Spectral Airglow
Temperature Imager, SATI,][]{wiens}.  They have also been used to
image gravity waves directly with narrow and broad filters
\citep[eg][]{taylor,garcia,li}.  These applications have been most
often used in concert with radar, lidar, balloons, rocketry, and
space-based experiments to achieve fully three-dimensional dynamical
information of O$_2$ and OH atmospheric layers.  None of these
observations ventured far into the red of today's deep depletion CCDs,
nor did the analyses achieve calibrated photometry.
Existing aeronomic data are therefore of limited use to LSST, as LSST
requires prior knowledge not only of the variability but also of the
expected absolute background level in the $y$ channel.  Our instrument
and data reduction techniques aim to provide all needed information at
once.

\subsection{Previous Astronomical Measurements}
\label{sec:astro}

Previous astronomical work is also insufficient, as the $y$ band sky
has not been directly characterized.  Extrapolating to the $y$ band
using other astronomical filters is problematic because of the highly
structured nature of the Meinel band emission (Figure
\ref{fig:Pixis_QE}, bottom panel).  Small changes in the filter's red
edge may suddenly alter the background level.  For example, the recent
UKIRT survey \citep{Warren07} does include a $Y$ band, but using an
infrared detector and filter whose total response function do not
match that expected for LSST.  \citet{Warren07} in fact showed that
their $Z$ band sky brightness from the UKIRT observations on Mauna Kea
was $0.4\;\mathrm{mag}\;\mathrm{arcsec}^{-2}$ brighter than that
obtained by SDSS in their $z$ band, at Apache Point, New Mexico.
Besides differences in effective filtering, the discrepancy may be
explained by differences in atmospheric conditions in time and in
space given the great geographical separation.
We therefore must assume that the
UKIRT median sky brightness of
$Y=17.4\;\mathrm{mag}\;\mathrm{arcsec}^{-2}$ (Vega) cannot be applied
to LSST.

Our effective filter function closely matches LSST's $y_3$ band and
thus minimizes potential errors due to differences in relative
spectral response; we additionally address the geographical problem by
observing at Cerro Tololo, a mountain ridge $10\,\mathrm{km}$
northwest of LSST's planned site, Cerro Pachon.
\cite{KK07} presented a historical overview of sky brightness
observations from Cerro Tololo, for the period 1992--2007. They
claimed that there was no discernable difference in sky brightness
between that site and Cerro Pachon. This leads us to conclude that our
$y$ band sky brightness measurements from Tololo should be applicable
to the LSST site on Pachon.


We conclude that previous work within aeronomy and astronomy is not
sufficient for characterizing the $y$ band sky background for LSST,
and that the precise nature of the chosen $y$ filter affects the
observed background due to the multitude of strong OH lines at $\sim
1\;\micron$ and redder.  The dependence of the apparent NIR sky
brightness on the filter and detector combination, and the potential
for variation between observatory sites, motivates the measurements
described here.

\section{Apparatus}
\label{sec:apparatus}

We used a Pixis system (detector, readout electronics, dewar and
shutter) from Princeton Instruments, equipped with a type 1024BR
CCD. This detector, made by EEV with their designation 47-10, was a
back-illuminated deep depletion sensor that exhibits enhanced
sensitivity at wavelengths near 900 nm due to the thicker region of
silicon from which photocharge is harvested.  The sensor was an array
of $1040 \times 1027$ pixels, each of which was $13.0\;\micron$ on a
side. The Pixis camera was equipped with a thermoelectric cooler that
maintained the detector at a temperature of $-20\;\mathrm{C}$. Dark
current was negligible compared to the flux seen from the sky.  We
measured the camera's gain to be $3.8\;\electrons\;\mathrm{ADU}^{-1}$.

Princeton Instruments designed the 1024BR CCD to suppress fringing
that would otherwise arise from night sky line emission.  We observed
no significant fixed fringing patterns in any of the images we
obtained, and did not have to apply any fringe-frame corrections to
the data.  We therefore judge any systematics in the photometry due to
fringing to be subdominant for the purposes of the results presented
here.

The Pixis camera was controlled from a laptop computer that ran the
Linux operating system. We used custom software to control the
exposure times and imaging sequence of the instrument. All images were
stored as 16 bit FITS files.
 
We attached a $17$--$55\;\mathrm{mm}$ variable focal length $f/2.8$
Canon EFS lens to the Pixis dewar, using a C-mount adapter made by
Birger Engineering.  The system was operated at $f/2.8$.  The $y$
filter we used was an interference filter manufactured to the LSST 
$y_2$ specification by Barr Associates, in a 
$2\;\mathrm{in} \times 2\;\mathrm{in}$ format. We
attached the filter to the front of the lens using a modified
professional photography filter holder that mated to the
$75\;\mathrm{mm}$ front threads of the Canon lens.  The lens's focal length
was adjusted so as to provide a plate scale of
$145\arcsec\;\mathrm{pixel}^{-1}$, and a square field of
view of $41\deg$.  Although we make use of a $y_2$ filter, the 
resultant system throughput more closely matches the LSST $y_3$ response
curve due to the difference in our detector QE and antireflection coatings on the optics from
the predicted LSST system.

The camera, lens and filter combination was mounted on a stationary
tripod. This arrangement did not provide pointing information in the
image headers, but as described below we were later able to determine
astrometry using the positions of cataloged sources.


Our optical configuration was not identical to that of LSST. The LSST
filters will be placed in an $f/1.23$ converging beam.  Light from a
point source therefore passes through the LSST filter in a (hollow)
cone with an opening half-angle between $14.2\deg$ and
$23.6\deg$. Obscuration from the secondary blocks rays at angles less
than $14.2\deg$.  The transmission curve through the interference
filter depends on the angle of incidence.  Rays that traverse the
filter at angles $\theta$ other than normal incidence produce a blue
shifted transmission function, shifted to the blue by roughly
\begin{equation}
  \delta \lambda = \lambda \left[\sqrt{1-n^{-1}\sin^2(\theta)} -1\right] 
\end{equation}
where $n$ is the effective index of refraction of the filter
dielectric.  The effective LSST passband is the appropriately weighted
integral of these angle-dependent transmission functions.

For our observations we placed the filter in front of the lens, so
rays from any point source pass through the filter as parallel
rays. The center of our field is observed through the filter at normal
incidence. The edges of our field are observed through the filter at
an incidence angle of $20\deg$. While we can't replicate the {\it
  distribution} of rays that will pass through the LSST filters, by
judiciously choosing the region in our field where we measure the sky
brightness, we can try to match the median LSST ray that traverses the
filter at an angle of $18.9\deg$ (LSST09). We therefore expect, even
for a perfect match between our interference filter and that of LSST,
slight differences in the effective passbands.  See Figure
\ref{fig:Pixis_QE} for a comparison of passbands.

\section{Observations}
\label{sec:observations}

We obtained images over four nights, UT 2007 Sep 13, 14, 18, and 19.
On the 13th we acquired data on the second half of the night, and
pointed the camera on the local meridian and low to the horizon at a
zenith angle of about $60\deg$.  For the remaining nights we aimed the
camera on the local meridian toward the equatorial, at approximately
$30\deg$ north.

The moon was 6\% illuminated on the first two nights and set at
$\hrmin{01}{43}$ UT.  The moon was 44\% illuminated on the last two
nights and rose at $\hrmin{05}{30}$ UT.  For the all observations, the
moon was not a complicating factor.  We estimate that the conditions
on the first two nights were photometric, whereas high clouds began to
form on the second half of the 18th and all of the 19th.  Table
\ref{tab:sod} summarizes the observations.

\begin{deluxetable}{cccc}
\tabletypesize{\footnotesize} 

  \tablecaption{Summary of the Data.\label{tab:sod}}

  \tablehead{\colhead{Night\tablenotemark{a}} & \multicolumn{2}{c}{Num
      exposures} &
    \colhead{Conditions\tablenotemark{b}} \\
    \colhead{~} & \colhead{$5\;\mathrm{min}$} &
    \colhead{$20\;\mathrm{s}$} & \colhead{~} }

  \tablewidth{0pt}

  \startdata

  2007 09 13 & 54  & 31 & Photometric \\
  2007 09 14 & 111 & 58 & Photometric \\
  2007 09 18 & 76  & 40 & First half of night photometric \\
  2007 09 19 & 67  & 35 & Clouds 

  \enddata

  \tablenotetext{a}{~Local date at start of night. }

  \tablenotetext{b}{~Based on our observations, archived SASCA
    red-channel animations (see
    \url{http://www.ctio.noao.edu/~sasca/sasca.htm}), GOES East
    10 micron satellite imagery (see
    \url{http://www.ctio.noao.edu/sitetests/GOES/}), CTIO DIMM and
    flux monitor data, and Blanco Cosmology Survey observers' logs. }

\end{deluxetable}%

On each night we elected to take all images at a fixed azimuth and
elevation in order to simplify our analysis of the temporal variation
in sky brightness. The nominal pointing changed somewhat from night to
night, however, because we disassembled the apparatus each morning.
We did not track the camera in R.A.  We alternated between sequences
of five $20\;\mathrm{s}$ (``short'') exposures and ten
$300\;\mathrm{s}$ (``long'') exposures throughout the night.  On
the 14th we acquired a total of 55 short exposures and 110 long
exposures.  The short frames had only slight trailing of stars, and
were useful for astrometry and photometric normalization using the
normal toolkit of astronomical data reduction for point sources. The
longer images offered better signal to noise for the sky brightness,
but stars trailed significantly in the EW direction.

\section{Data Reduction}
\label{sec:reduction}







\begin{figure}
\plottwo{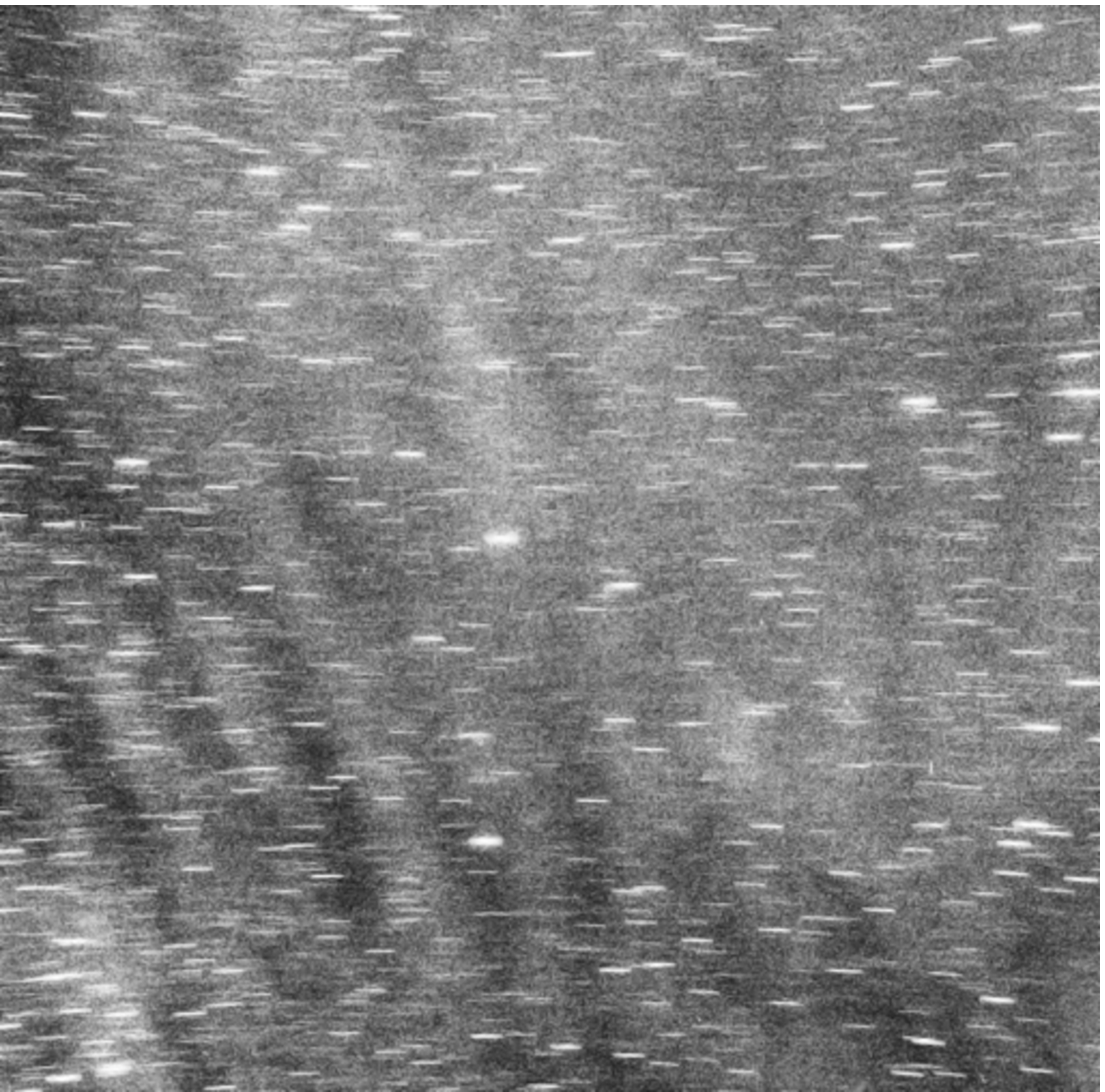}{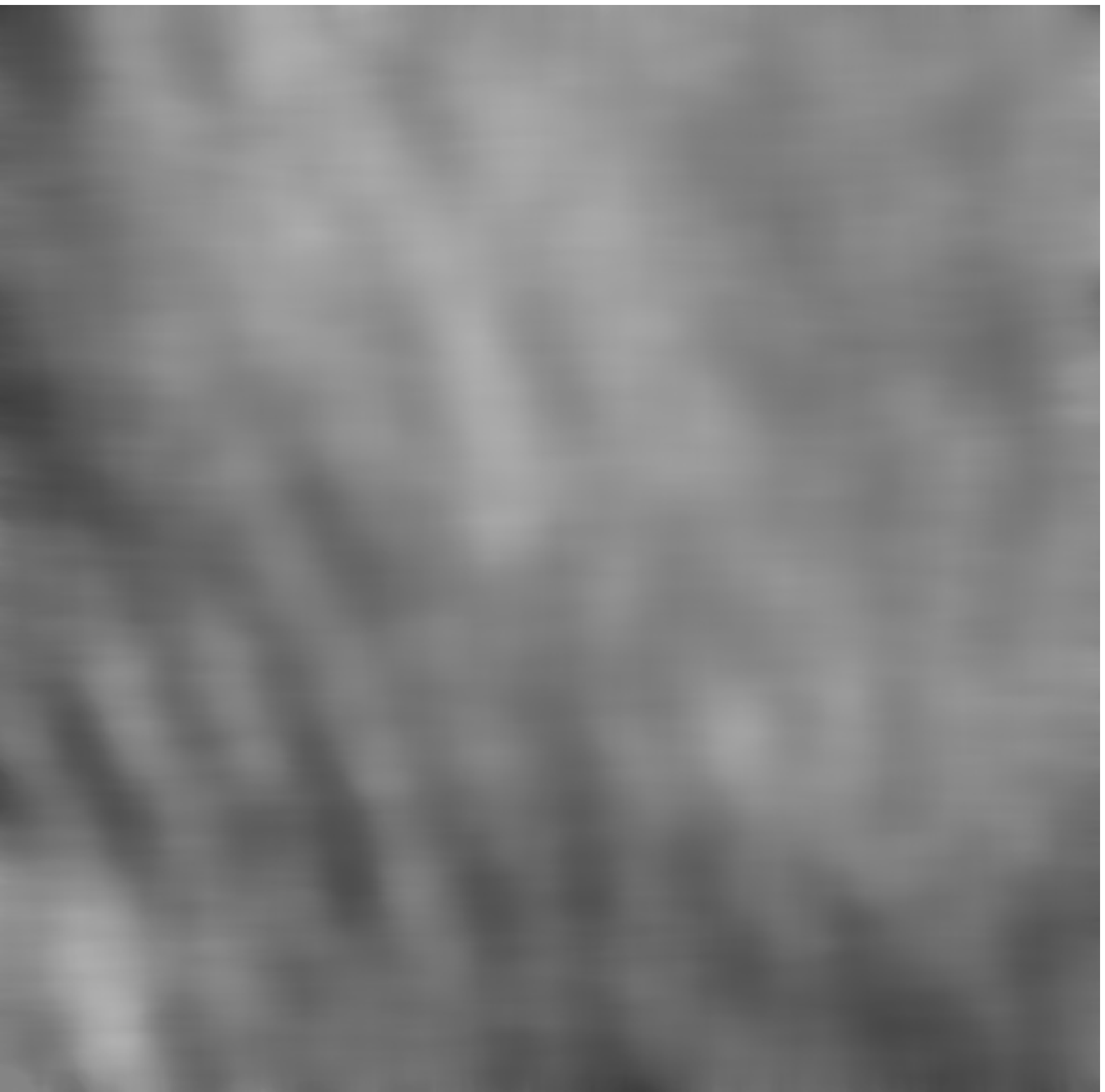}

\caption{A flat-fielded $300\;\mathrm{s}$ exposure in $y_3$, taken
  at fixed altitude and azimuth (righthand frame has stars removed to
  emphasize the airglow structure). The star trails subtend about
  $1.25 \deg$ in the EW direction; north is up and east is right. This
  image shows two distinct gravity waves traced by NIR narrowband
  emission, which gives rise to short term variability in $y_3$ band
  sky brightness.  The high frequency ripples have wavelength of about
  5 deg and propagate at between zero and about a degree per minute.
  The low frequency ripple's wavelength is roughly the field of view
  (40 deg) and travels at a degree per minute.  Root mean square sky
  variations in a given frame are up to 4\%, while the overall sky
  level over a full night varies by a factor of 2.}
 
\label{fig:long}
\end{figure}

\begin{figure}
\plottwo{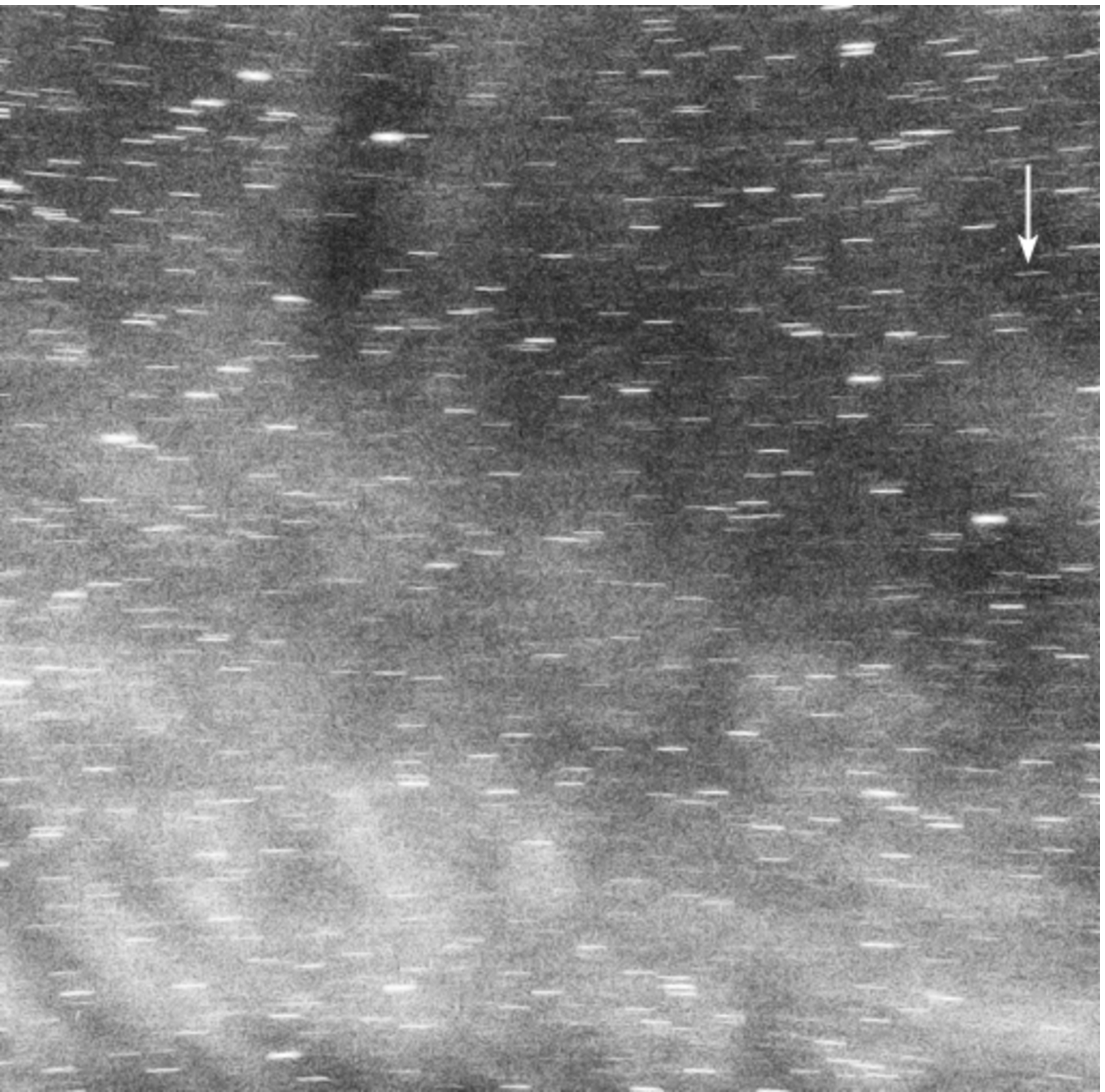}{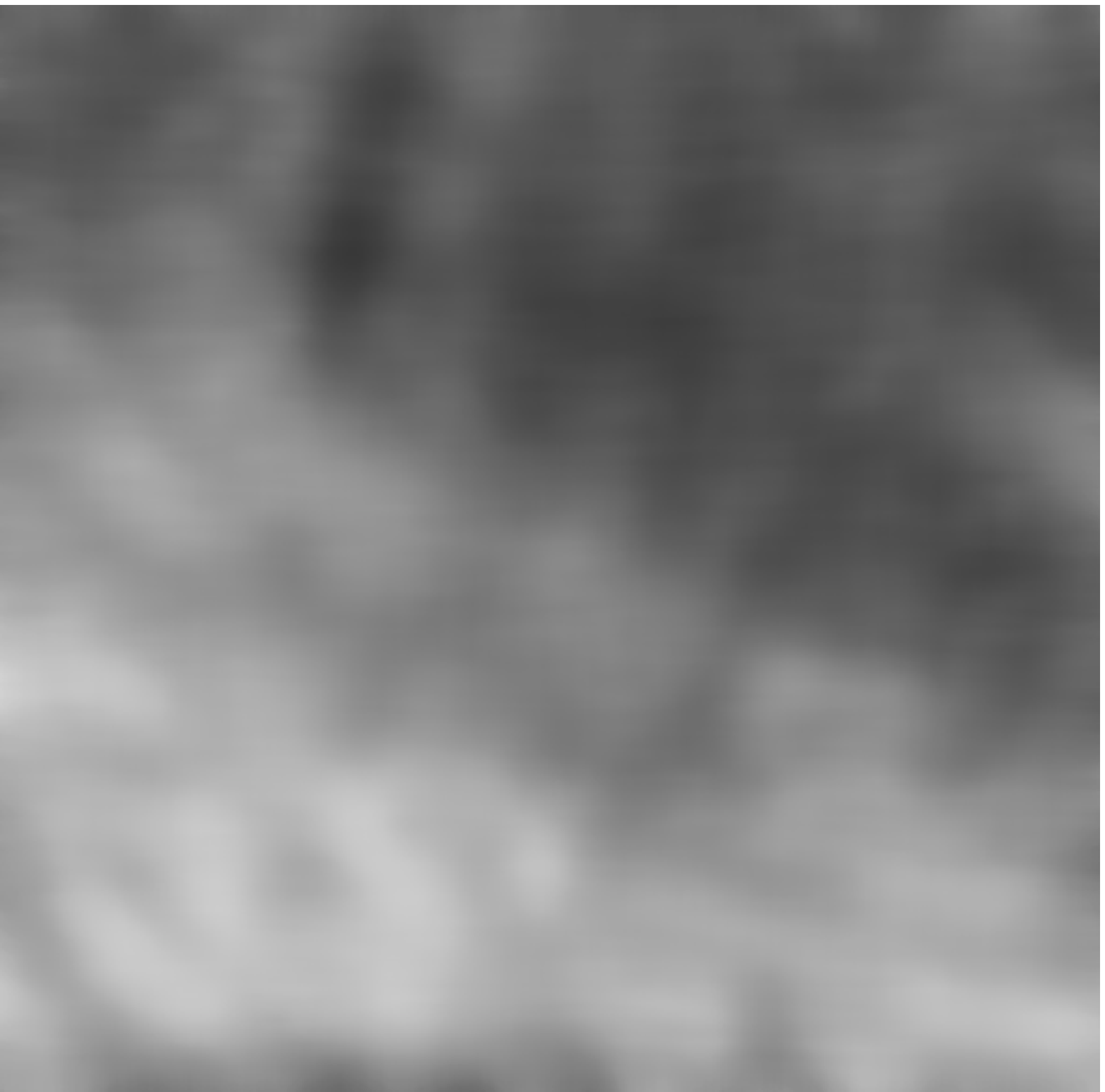}

\caption{Same as Figure \ref{fig:long}, but taken later in the night,
  just as our calibration star $\rho$ Ceti entered the field of view
  (upper right corner of lefthand frame).}
 
\label{fig:long2}
\end{figure}

We used the utilities in IRAF to do a line-by-line bias subtraction of
each of the images, using the overscan.  The $300\;\mathrm{s}$
exposures were combined into a normalized sky flat, by taking a
flux-scaled median of the frames. This normalized sky flat was then
used to flat-field each of the images. The normalized sky flat had a
fractional response suppression of about 20\% from the center to the
corners of the field, presumably due to vignetting and perhaps plate
scale variation from the lens.  This variation in flux sets an upper
bound for potential systematic errors in sky brightness due to
passband dependence on the angle of incidence, at
$0.2\;\mathrm{mag}$. We take this as an extreme upper limit, since we
do expect $\cos^n$ factors to roll off the response at the edges of
the field.  Because we are measuring sky brightness, the stacked sky
flat is indeed appropriate for correcting the pixel-to-pixel
variations on the scale of a PSF.  
Our $41\deg\times41\deg$ field of
view rendered twilight flat-fielding impossible due to sky gradients
over such a large field and to the short window of time over which
useful sky levels per pixel could be acquired, and we had no system in
place for obtaining the equivalent of dome-flats.  
As noted in \S
\ref{sec:apparatus}, the 1024BR CCD was specifically designed to
suppress fringing, and indeed we observed no fringe patterns in any of
our data and therefore made no fringe-frame corrections.  
Figures
\ref{fig:long} and \ref{fig:long2} show typical long exposures, after
flat-fielding and bias subtraction.


 

\section{Results}
\label{sec:results}

We first explored relative measurements of the sky background, because
we expect relative errors to be significantly smaller than the overall
uncertainty from absolute zeropoint calibration.  We then estimated
the overall zeropoint in the Vega and AB magnitude systems.

\subsection{Spatial Variability}
\label{sec:spatial}

The gravity waves we observed in the vast majority of our images are
easily seen in Figure \ref{fig:long} and \ref{fig:long2}.  These
structures were not due to flatfielding errors.  This can be deduced
from the two example images themselves, which were taken from the same
night and had the same flatfield correction applied to them.  Roughly
speaking, the locations of large-scale brightness minima in Figure
\ref{fig:long} correspond to maxima in Figure \ref{fig:long2}.
Indeed, these waves appear and disappear, and move coherently in
images from any given night viewed consecutively.  Their direction of
propagation is perpendicular to their wavefronts.

To quantitatively estimate the effect of the waves on sky brightness
at any given time, we computed the mode of the sky brightness values
for three regions of the detector, for each of the flatfielded 300~s
images.  The reference region was the central $500\times500$ pixel
area of the detector.  This was used to estimate the zero-level over
the large area of sky that we observed.  Our ``calibrator'' region was
a smaller, circular aperture of radius $50$ pixels near the detector
corner that we ultimately used for zeropoint calibration
(\S~\ref{sec:absolute}).  The ``control'' region was another 50 pixel
radius aperture near the opposite corner, but at the same angle of
incidence with respect to the camera boresight.  The calibrator and
control regions were chose to be smaller than the size of the smallest
coherent spatial gravity waves in the images, whereas the reference
region was chosen to be much larger.

\begin{figure*}
  \plotone{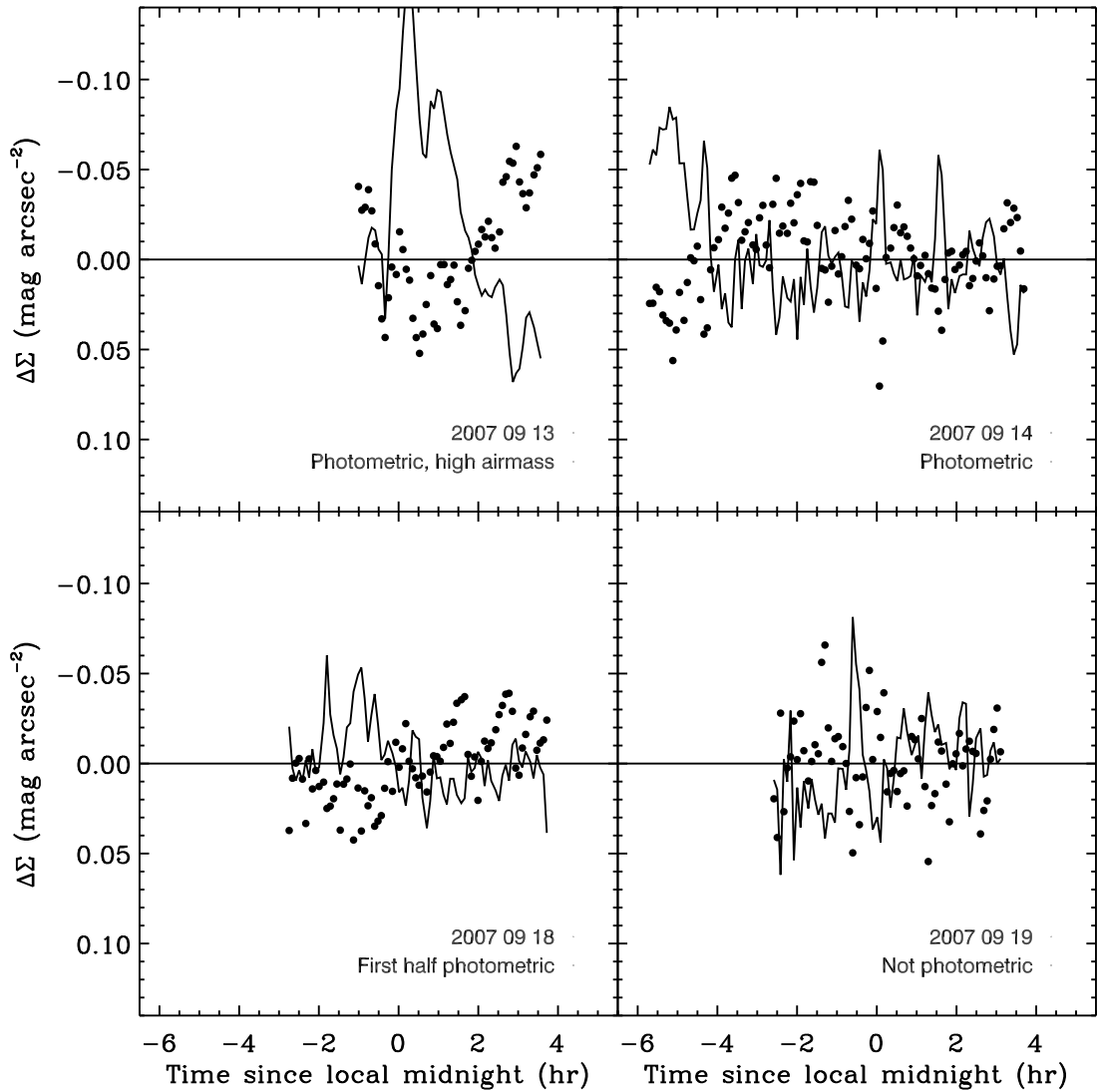}
  \caption{Spatial variation in $y_3$ band sky brightness versus time,
    showing $2\%$--$5\%$ rms variability in flux. This plot
    shows the fractional difference in sky brightness between our
    calibrator region (circular aperture of radius $50$ pixels) versus
    a central reference region of size $500\times500$ pixels (points),
    and between a second $50$ pixel radius region on the opposite
    corner of the chip and the reference region (line).  The large
    variation on the 13th is due to the extreme difference in airmass
    between to two corner regions: 5.8 for the northeast corner and
    1.3 for the southwest corner.  Early on in the night on the 14th,
    the Galactic plane fell within the NW aperture, again causing
    larger than typical variation.}
  \label{fig:spatial}
\end{figure*}

We calculated the mode of pixel values in each aperture to attain sky
counts $c$ in ADU pixel$^{-1}$.  Relative magnitudes between each
corner region and the central reference region were calculated as
$\Delta\Sigma=2.5\log(c_{\mathrm{cal}}/c_{\mathrm{ref}})$, (likewise
for $c_{\mathrm{ctrl}}/c_{\mathrm{ref}}$) with units of
mag~arcsec$^{-2}$, assuming a common pixel scale between the two
apertures.  Figure \ref{fig:spatial} shows these fractional
differences over each night.

Observations on the night of the 13th were taken at $\sec(z)=2$
airmasses at the center of the detector, with the calibrator region at
$1.4$ airmasses and the control region at approximately $4$ airmasses,
due to our large field of view.  This gave rise to rms
spatial sky brightness variations of $3\%$ between the calibrator and
the reference region, $6\%$ between the control and the reference
region, and $8\%$ between the calibrator and control regions.  On the
other three nights, the camera pointed at about $1.2$ air masses, with
the calibrator region at $1.0$ air masses and the control region at
$1.4$ air masses.  Ignoring the early part of the 14th, where the
Galactic plane fell within the control region, these nights exhibited
2\%--3\% rms spatial variability with respect to the reference, and
3\%--4\% rms variability between the calibrator and control regions.  We
conclude that, on angular scales of the wavelength of the gravity waves, $\gtrsim2\%$, the spatial variability in the $y$ band is typically
3\%--4\% rms.  Our relative photometric errors are
expected to be subdominant to this effect.  Under the assumptions that
gain, exposure time, and pixel scale are all constant in any given
image, these results are independent of those quantities and of any
absolute flux calibration.



\subsection{Temporal Variability}
\label{sec:temporal}

\begin{figure*}
\plotone{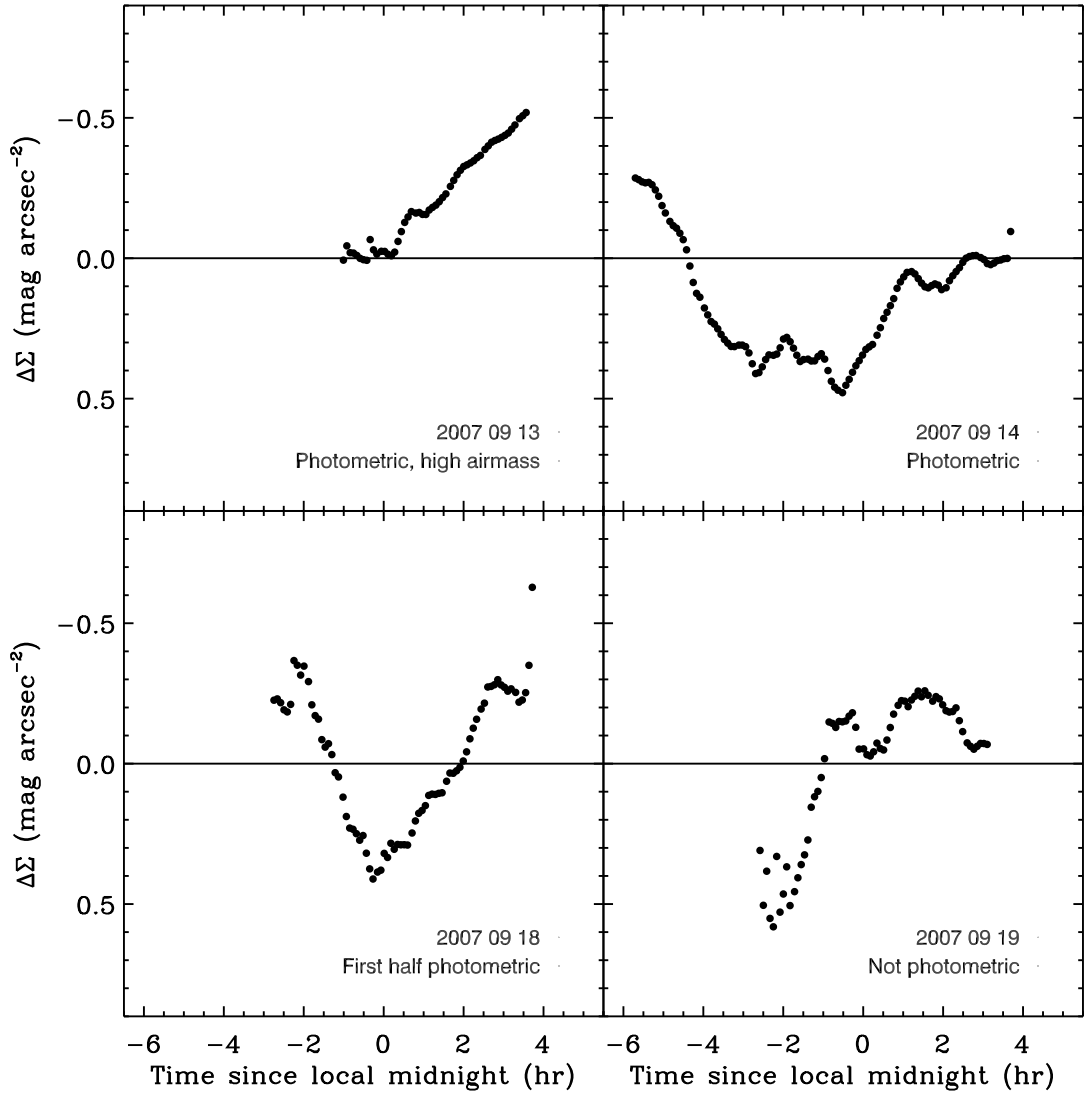}

\caption{ Temporal variation in $y_3$ band sky brightness versus time,
  showing highly coherent variability of up to a factor of 2 in flux
  over the course of a night.  Flux was taken along a fixed line of
  sight that is $20\deg$ from the zenith and directed north. Data from
  the 13 and 14 are most probably photometric, the first half of the
  18 may be useful, and the 19 is not photometric.  The sky brightness
  levels in this plot were obtained from the same $2.13\deg$ radius
  circular region on the detector, and constitutes a differential
  measurement with very high signal to noise ratio.  The fluctuations
  on both long and short timescales are real. We consider the
  Vega-based $y_3$ band zeropoint to be reliable to
  $0.2\;\mathrm{mag}$, but this systematic uncertainty is swamped by
  the temporal variability in sky brightness.  To convert to AB
  $\mathrm{mag}\;\mathrm{arcsec}^{-2}$, add $0.5\;\mathrm{mag}$ to the
  $y^{\mathrm{Vega}}_2$ values.}
 
\label{fig:sky}
\end{figure*}

The overall temporal sky variability is estimated using a similar
approach as in the previous section.  The fluxes within the
$500\times500$ pixel reference region are compared to the mean over
all 4 nights, and plotted in Figure \ref{fig:sky}.  Here we are
comparing $2.5$ times the log of sky counts in ADU pixel$^{-1}$ {\it
  between} images rather than within them, but because all exposure
times, gain values, and pixel regions (and therefore pixel scales)
were the same, this is equivalent to measuring relative sky brightness
in mag arcsec$^{-2}$.  Again, all that needs to be done to bring this
to an absolute scale is to add a zeropoint in mag arcsec$^{-2}$.

Immediately evident both in the Figure and in the raw images viewed in
succession is a significant overall sky level variability.  The
peak-to-peak variation on the 13th is 1.8 in flux, and on the 14th,
18th, and 19th are 2.0, 2.7, and 2.2 in flux, respectively.  We have
used a huge region of about $400$ deg$^2$, and we expect our
statistical uncertainties from both the gravity waves and from
electron counting to be subdominant to the temporal variability we
measure.  Changes in overall sky level are highly correlated in time,
and all of the structure seen in Figure \ref{fig:sky} is due to the
atmospheric emission variablity.

The relative results have made use of only ADU counts, assuming
constancy for all other quantities that are needed for an absolute
flux calibration.  Our final task is to estimate the overall
zeropoint.

\subsection{Absolute Sky Brightness}
\label{sec:absolute}

In order to convert measurements from $\mathrm{ADU}\;
\mathrm{s}^{-1}\; \mathrm{pixel}^{-1}$ to $\mathrm{mag}\;
\mathrm{arcsec}^{-2}$ we determined the gain, pixel scale, and
magnitude zeropoint.  We measured the gain to be
$3.8\;\electrons\;\mathrm{ADU}^{-1}$.  The pixel scale and zeropoint
measurements were both accomplished using a standard star.

We used photometry of a type A0V standard star to establish a
conversion between our instrumental magnitudes and a Vega-based $y_3$
band magnitude scale. This follows the approach used in
\cite{Hillen02} and \cite{Hew06}, where A0V stars are used as proxies
for Vega to establish a zeropoint on the Vega magnitude scale.

We were fortunate to have beta-test access to the
astrometry.net web utility (\url{http://astrometry.net/}), which has
the remarkable ability to determine a WCS solution for an arbitrary
image. We input one of our images, frame 110 (taken at UT
$\ra{06}{19}{22}$) to this site and learned that the image was
centered at $\mathrm{R.A.} = \ra{01}{17}{34.8}$, $\mathrm{Decl.} =
\dec{-02}{20}{49}$ (J2000). We then queried the SIMBAD archive to find
A0 stars that coincided with this field. The A0V star $\rho$ Ceti at
$\mathrm{R.A.} = \ra{02}{25}{57.0}$, $\mathrm{Decl.} =
\dec{-12}{17}{26}$ (J2000) (also designated as HR 708 and SAO 148385)
met our criteria. This object has cataloged magnitudes that are listed
in Table \ref{rhoCet}.

\begin{deluxetable}{ccc}
\tabletypesize{\footnotesize} 

 \tablecaption{$\rho$ Ceti data from the literature.\label{rhoCet}}

 \tablehead{
   \colhead{Band\tablenotemark{a}} & 
   \colhead{$m^{\mathrm{Vega}}$} &
    \colhead{$\sigma_m$}}

  \tablewidth{0pt}

  \startdata

  $U$ & 4.85 & 0.03 \\
  $B$  & 4.85 & 0.03\\
  $V$ & 4.88 & 0.03\\
  $J$ & 5.08 & 0.20 \\
  $H$ & 4.86 & 0.02\\
  $K$ & 4.81 & 0.02

  \enddata

  \tablenotetext{a}{~$UBV$ data are from \citet{UBV1} and $JHK$ data
    are from \citet{2Mass}.}

\end{deluxetable}%

We identified $\rho$ Ceti in five of the short exposure images, and
used IRAF to perform aperture photometry. The resulting centroid
positions, object fluxes, and sky background values are given in Table
\ref{rhoCetData}.  We determined the flux from the star by averaging
these measurements, obtaining $119 \pm 17 \; \mathrm{ADU}$ accumulated
in $20\; \mathrm{s}$, or
$5.95\pm0.85\;\mathrm{ADU}\;\mathrm{s}^{-1}$ for an object with a
Vega-based magnitude of $y^{\mathrm{Vega}} = 4.86\;\mathrm{mag}$.
This fractional flux uncertainty corresponds to a systematic
uncertainty in the Vega-based magnitude zeropoint of
$0.15\;\mathrm{mag}$.  Using the long exposures would not gain much in
signal to noise ratio, because the SNR is independent of exposure time
when an object trails in an image.

\begin{deluxetable}{cccccc}
\tabletypesize{\footnotesize} 

  \tablecaption{$\rho$ Ceti data.\label{rhoCetData}}

  \tablehead{
    \colhead{Frame} & 
    \colhead{UT} &
    \colhead{$x$\tablenotemark{a}} & 
    \colhead{$y$\tablenotemark{a}} &
    \colhead{$\phi$\tablenotemark{b}} & 
    \colhead{$c$\tablenotemark{c}} \\
    \colhead{~} & 
    \colhead{~} & 
    \colhead{(pixel)} & 
    \colhead{(pixel)} & 
    \colhead{(ADU)} &
    \colhead{($\mathrm{ADU}\;\mathrm{pixel}^{-1}$)} }

  \tablewidth{0pt}

  \startdata

  110 & $\ra{06}{19}{01}$ & 925.4 & 257.7 & ~92.6 & 14.6  \\
  111 & $\ra{06}{19}{22}$ & 923.5 & 257.8 & 117.0 & 15.5 \\
  112 & $\ra{06}{19}{44}$ & 921.7 & 257.9 & 129.6 & 13.6 \\
  113 & $\ra{06}{20}{05}$ & 919.1 & 257.8 & 119.2 & 15.3 \\
  114 & $\ra{06}{20}{26}$ & 916.8 & 258.0 & 136.7 & 15.3

  \enddata

  \tablenotetext{a}{~Centroid of the star.}
  \tablenotetext{b}{~$\rho$ Ceti flux.}
  \tablenotetext{c}{~Local sky.}

\end{deluxetable}%

The rate of motion of a star at a known right ascension and
declination allows for the mapping from pixels to arcseconds in the
region near the star.  The data from Table \ref{rhoCetData} were also
used to this end.  $\rho$ Ceti was at a declination of $-12.29\;\deg$,
implying an apparent motion across the sky of $\mu =
15\arcsec\;\mathrm{s}^{-1}\;\cos{12.29} =
14.66\arcsec\;\mathrm{s}^{-1}$.  We detected the object as
moving $8.6\;\mathrm{pixel}$ in $85\;\mathrm{s}$, or
$0.101\;\mathrm{pixel}\;\mathrm{s}^{-1}$. This implies a plate scale
of $144.9\arcsec\;\mathrm{pixel}^{-1}$, so that each pixel
subtends an area of $20,995\;\mathrm{arcsec}^2$, or
$5.8\;\mathrm{arcmin}^2$.  We consider this plate scale to be
determined with a fractional uncertainty of about $5\%$, due to both
the $1\;\mathrm{s}$ granularity in the timing between exposures and
the centroid measurement uncertainty for the star.  Our dynamic plate
scale determination is also within $6\%$ of the
$153.4\arcsec\;\mathrm{pixel}^{-1}$ that was returned by
astrometry.net, based on matching sources across the field to
cataloged objects. In one of our longer exposures we measured the
trail length of a star at this same location on the array as spanning
$31.2\;\mathrm{pixels}$ in a $300\;\mathrm{s}$ exposure, again
confirming the plate scale determination at the $5\%$ level.

The calibration source, and the region around it where we determined
the sky brightness values, reside near a corner of the field. The rays
that are brought to a focus at this location pass through the filter
at an angle of $18.6\deg$, while the median LSST incidence angle is
$18.9\deg$. We therefore consider the center of our effective passband
at this location to be a good match to LSST's, but the shape at the
passband edges will still differ somewhat.

With the photometric calibration and plate scale in hand, we were in a
position to convert the sky brightness values, $\Sigma$, into
Vega-based units of $\mathrm{mag}\;\mathrm{arcsec}^{-2}$. We converted
the sky brightness values into units of
$\mathrm{ADU}\;\mathrm{s}^{-1}\;\mathrm{arcsec}^{-2}$ by dividing by
20,995 $\mathrm{arcsec}^2\;\mathrm{pixel}^{-1}$.  We then computed
the Vega-based sky brightness using the flux from the comparison star,
as
\begin{equation}
  \Sigma^{\mathrm{Vega}}=-2.5\log(\phi/5.95) + 4.85\;\frac{\mathrm{mag}}{\mathrm{arcsec}^2}
\end{equation}
where $\phi$ is the sky flux in
$\mathrm{ADU}\;\mathrm{s}^{-1}\;\mathrm{arcsec}^{-2}$.  Note that we
selected a region for sky brightness determination that surrounded the
calibration star to minimize our sensitivity to large scale
flatfielding errors and pixel scale variations.  The final zeropoint
that converts all temporal results of Figure \ref{fig:sky} to Vega
magnitudes is $\Sigma = \Delta\Sigma
+17.8\;\mathrm{mag}\;\mathrm{arcsec}^{-2}$ (Vega).

We also supply the conversion of Vega magnitudes through the $y$ band
to AB magnitudes, because, while the AB and Vega magnitude scales are
matched at the V band, they monotonically diverge at longer
wavelengths.  The Spitzer flux conversion utility\footnote{At
\url{http://ssc.spitzer.caltech.edu/warmmission/propkit/pet/magtojy/}} delivers
$\mathrm{AB}^{\mathrm{Vega}}_{y}=0.5\;\mathrm{mag}$ with an
uncertainty of $0.05\;\mathrm{mag}$,, where
$y^{\mathrm{AB}}=y^{\mathrm{Vega}}+\mathrm{AB}^{\mathrm{Vega}}_{y}$.
This level of uncertainty is adequate for our purposes here. The
conversion also agrees within the uncertainties with the AB-to-Vega
offset terms presented in Table 7 of \cite{Hew06}.  Thus the
conversion $\Sigma = \Delta\Sigma +18.3\;\mathrm{mag}$ (AB) brings all
temporal results of Figure \ref{fig:sky} into the AB system.  Table
\ref{systematics} presents a summary of the potential sources of
systematic error that might afflict these results.

\begin{deluxetable}{ll}

 \tablecaption{Systematic error budget.\label{systematics}}

  \tablehead{
    \colhead{Source} & 
    \colhead{Error\tablenotemark{a} (mag)} } 

  \tablewidth{0pt}

  \startdata

  Sky area pixel$^{-1}$ &  0.05   \\
  Vega-system zeropoint & 0.15  \\
  AB system zeropoint & 0.05  \\
  Stray \& scattered light & Unknown  \\
  Photometric conditions & 0.01  \\
  Passband errors & Unknown

  \enddata

  \tablenotetext{a}{~Estimated upper limit of systematic error.}

\end{deluxetable}%

\section{Conclusions}
\label{sec:conclusions}

We have performed first measurements of $y_3$ band sky brightness in
Chile from Cerro Tololo, and have argued that these results to apply
to the nearby LSST site on Cerro Pachon.  We observed rms spatial
variability of 3--4\% due to atmospheric gravity waves on scales of a
few degrees and larger, and a factor of 2 variability in flux over the
course of the nights we observed.  These relative results make no reference
to absolute calibrations, and we argue that they apply to nearly any
$y$ band variant because the sky background is overwhelmingly due to
narrowband OH emission.  Finally, we estimate an absolute zeropoint
and find the mean sky level is about
$17.8\;\mathrm{mag}\;\mathrm{arcsec}^{-2}$.  Our absolute calibration
suffers up to $20\%$ systematic uncertainty from zeropoint and pixel
scale measurement errors, while our relative photometry is uncertain
at about a percent.

Previous aeronomy work has characterized the OH background in depth,
and suggests that the variability is somewhat predictable. We expect
the sky background through any similar, CCD-based $y$ band variant to
fluctuate significantly, although nominal variation and absolute
levels will differ depending on the precise nature of the total filter
response curve.  Our qualitative results and conclusions should apply
to any $y$ band.

The variability suggests that sky survey projects would benefit from a
$y$ band sky brightness monitor, and an adaptive strategy that would
exploit times when the $y$ sky is dark to take observations in this
band.  The background in $y$ appears uniform to within $4\%$ over much
of the sky, so we advise it is better to chase {\it times} of low
background than to find {\it directions} of dark sky in $y$.  The
coherent spatial structures in $y$ band airglow present a potential
difficulty to sky-based illumination correction strategies.  Assuming
random Gaussian fluctuations, 15 images or more are needed to achieve
sub-percent uniformity over fields of view larger than about a degree.
The spatial coherence of the gravity waves means the fluctuations are
not random, making this a lower limit.

We have demonstrated the effectiveness of this apparatus. Given the
observed variability in $y_3$ band sky brightness, there is
considerable merit in implementing a long term monitor at the LSST
site. Having long term sky brightness data in hand early will inform
the optimal scheduling of the LSST system.

\acknowledgments

We thank the LSST Corporation, Harvard University and the Department
of Energy Office of Science and the National Science Foundation for
their support.  The LSST design and development activity is supported
by the National Science Foundation under Scientific Program Order
No. 9 (AST-0551161) through Cooperative Agreement AST-0132798.
Additional funding
comes from private donations, in-kind support at Department of Energy
laboratories and other LSSTC Institutional Members.  We thank the CTIO
scientific and technical staff for their invaluable help in setting up
these observations.  We are also very grateful to the team that is
establishing the astrometry.net (\url{http://astrometry.net/}) online
resource, which we used in its testing phase to determine the centers
of the images we obtained. The authors gratefully acknowledge the
referee for insightful comments and recommendations.  This publication
makes use of data products 
from the Two Micron All Sky Survey, which is a joint project of the
University of Massachusetts and the Infrared Processing and Analysis
Center/California Institute of Technology, funded by the National
Aeronautics and Space Administration and the National Science
Foundation.

\end{document}